\documentclass[12pt]{article}
\usepackage{graphicx,subfigure,pstricks,pst-node,psfrag,amsthm,amssymb,amsmath,dsfont}
\usepackage[round]{natbib}

\usepackage[normalem]{ulem}

\newtheorem{theorem}{Theorem}[section]

\newcommand{\bs}{\mbox{\boldmath $s$}}

\newcommand{\bw}{\mbox{\boldmath $w$}}
\newcommand{\bx}{\mbox{\boldmath $x$}}
\newcommand{\by}{\mbox{\boldmath $y$}}
\newcommand{\bz}{\mbox{\boldmath $z$}}

\newcommand{\bA}{\mbox{\boldmath $A$}}

\newcommand{\bF}{\mbox{\boldmath $F$}}

\newcommand{\bI}{\mbox{\boldmath $I$}}
\newcommand{\bK}{\mbox{\boldmath $K$}}
\newcommand{\bM}{\mbox{\boldmath $M$}}

\newcommand{\bS}{\mbox{\boldmath $S$}}
\newcommand{\bU}{\mbox{\boldmath $U$}}
\newcommand{\bV}{\mbox{\boldmath $V$}}
\newcommand{\bW}{\mbox{\boldmath $W$}}

\newcommand{\bX}{\mbox{\boldmath $X$}}
\newcommand{\bZ}{\mbox{\boldmath $Z$}}

\title{
\vspace{-1in}
{\bf Homotopic Group ICA for Multi-Subject Brain Imaging Data}
\vspace{-0.2in}}
\author{Juemin Yang, Ani Eloyan, Anita Barber, 
Mary Beth Nebel, \\ Stewart Mostofsky, James J. Pekar,  Ciprian  Crainiceanu and Brian Caffo } 
\vspace{-0.2in}

\date{}

\begin{document}\maketitle
\vspace{-0.4in}
\begin{abstract} {\footnotesize Independent Component Analysis (ICA)
    is a computational technique for revealing latent factors that
    underlie sets of measurements or signals. It has become a standard
    technique in functional neuroimaging. In functional neuroimaging,
    so called group ICA (gICA) seeks to identify and quantify networks
    of correlated regions across subjects.  This paper reports on the
    development of a new group ICA approach, Homotopic Group ICA
    (H-gICA), for blind source separation of resting state functional
    magnetic resonance imaging (fMRI) data.  Resting state brain
    functional homotopy is the similarity of spontaneous fluctuations
    between bilaterally symmetrically opposing regions (i.e. those
    symmetric with respect to the mid-sagittal plane)
    \citep{zuo2010growing}.  The approach we proposed improves network
    estimates by leveraging this known brain functional homotopy.
    H-gICA increases the potential for network discovery, effectively
    by averaging information across hemispheres.  It is theoretically
    proven to be identical to standard group ICA when the true sources
    are both perfectly homotopic and noise-free, while simulation
    studies and data explorations demonstrate its benefits in the
    presence of noise. Moreover, compared to commonly applied group
    ICA algorithms, the structure of the H-gICA input data leads to
    significant improvement in computational efficiency. A simulation
    study comfirms its effectiveness in homotopic, non-homotopic and
    mixed settings, as well as on the landmark ADHD-200 dataset. From
    a relatively small subset of data, several brain networks were
    found including: the visual, the default mode and auditory
    networks, as well as others. These were shown to be more
    contiguous and clearly delineated than the corresponding ordinary
    group ICA. Finally, in addition to improving network estimation,
    H-gICA facilitates the investigation of functional homotopy via
    ICA-based networks.}
\end{abstract}


\newpage

\section{Introduction}\label{s:intro}
The function of the human brain during rest can be investigated using
various functional measurement techniques
\citep{biswal1995functional,gusnard2001searching,gao2009evidence}.  Patterns of
resting-state brain functional across individuals provide insights
into baseline activity of the human brain in the absence of
experimental stimuli.

Resting state functional magnetic resonance imaging (rs-fMRI),
obtained using blood oxygen level dependent (BOLD) signals, is a key
driving force for investigating baseline fluctuations in brain
activity.  Much recent attention has been focused on identifying the
possible origins and clinical manifestations of variation in the BOLD
signals from rs-fMRI data.  Neuroimaging studies have identified
associations between resting state brain networks estimated via fMRI
data with aging, cognitive function and neurological and psychiatric
disorders \citep{damoiseaux2008reduced,rombouts2005altered}.  One
popular approach for locating putative networks is blind source
separation, which decomposes neuroimaging data into an outer product
of spatial maps multiplied by their respective time courses. Notably,
blind source separation does not require a specific fMRI paradigm, so
is applicable to resting state data.  Two popular exploratory data
analysis techniques for blind source separation are Principal
Component Analysis (PCA) and Independent Component Analysis (ICA).
ICA can be distinguished from PCA by its focus on model-level
independence and non-Gaussianity. Moreover, ICA as a matrix
decomposition does not yield decomposition vectors that are
orthonormal. Finally, ICA is usually applied after PCA-based
dimensional reduction, and thus can be thought of as a non-orthonormal
reorganization of PCA.

ICA has become popular for analyzing neuroimaging data, having been
successfully applied to single-subject analysis
\citep{guo2008unified,beckmann2004probabilistic,mckeown1997analysis}.
The extension of ICA to group inferences provides common independent
components across subjects and enables identification of putative
underlying brain networks for the group. 
Several multi-subject ICA approaches have been proposed: \cite{calhoun2001method} presented
(and named) group ICA (gICA) and created the Matlab toolbox (GIFT), which
provides  estimation. GIFT consists of a first-dimension
reduction using PCA for each subject, followed by a temporal
concatenation of the reduced data, after which ICA is then applied to
the aggregated data.  More recently, \cite{beckmann2005tensorial}
proposed a tensor ICA (TICA) by extending the single-session
probabilistic ICA (PICA) \citep{beckmann2004probabilistic}. TICA
factors the multi-subject data as a tri-linear combination of three
outer products, which represents the different signals and artifacts
present in the data in terms of their temporal, spatial, and
subject-dependent variations. Other grouping methods proposed perform ICA on each subject and combine the output into a group by using self-organizing clustering \citep{esposito2005independent} or spatial correlation \citep{calhoun2001fmri}.

Among existing methods, GIFT is perhaps the most commonly used for performing
group ICA analysis of multi-subject fMRI data. The spatial
independence assumed by GIFT, and any other spatial ICA algorithms, is well 
suited to the spatial patterns seen in fMRI
\citep{mckeown1997analysis}. Moreover, its empirical performance has
been consistently validated
\citep{sorg2007selective,garrity2007aberrant,sambataro2010age}.

This paper describes a novel modified group ICA method - homotopic
group ICA (H-gICA) - to identify the underlying patterns of brain
activity for functionally homotopic putative networks.  Left to right bilaterally symmetric
patterns of interhemispheric synchrony and co-activation are among the most frequent
findings in neuroimaging studies \citep{toro2008functional}.
Functional homotopy, the
similarity of spontaneous fluctuations between bilaterally symmetrically opposing regions 
with respect to the mid-sagittal plane is a ``fundamental characteristic of the brain's
intrinsic functional architecture'' \citep{zuo2010growing}.  Via H-gICA,
the information of homotopic brain function is utilized to improve the
identification of underlying brain networks.  A spatial independence
assumption is made relating to all voxels in each hemisphere.  In a
simulation study, H-gICA is shown to be preferable to (our
implementation of) GIFT when the actual signals are homotopic and is
competitive with GIFT under non-homotopic signal conditions.  The
efficacy of H-gICA methodology is demonstrated by an application to
the ADHD-200 dataset \citep{milham2012open,eloyan2012automated}.  From
the fifteen components produced by H-gICA, several common brain
networks were found, being clearly represented in smoother, more clearly
delineated and contiguous volumes than ordinary gICA.  The main
networks found include the visual, default mode and auditory.
In addition to improving network estimation, H-gICA allows for the
investigation of functional homotopy via ICA-based networks. Here the
quantification of functional homotopy of such networks are defined
using a similar concept to that of \cite{joel2011relationship}. Specifically,
by having left and right hemispheric terms in the model, investigations
of ICA weight matrices includes homotopic information, unlike ordinary
group ICA.

The remainder of the paper is organized as follows. 
 ICA and  fast, fixed-point algorithm known as FastICA proposed by
\citep{hyvarinen1999fast} is reviewed in Section \ref{sec:background},
and subsequently used in the estimation of H-gICA. Section \ref{sec:methods}
is the theoretical body of the paper: Section \ref{sec:preprocessing} discusses about data preprocessing issues; Section \ref{sec:hgICAmodel}
introduces the H-gICA model; Section \ref{sec:connection}
proves that H-gICA and GIFT coincide under noise-free settings;
and Section \ref{sec:homotopy} provides a measure of the functional
homotopy for ICA-based networks.  Section \ref{sec:simulation}
provides a simulation study to demonstrate the effectiveness of H-gICA
under homotopic, non-homotopic and mixed settings. This allows a
comparison versus GIFT, demonstrating that by using intrinsic
 functional homotopy of the brain, H-gICA improves the power of locating the
underlying brain networks. Section \ref{sec:application} provides the
application of the H-gICA on the ADHD-200 dataset, an open source
dataset with 776 children and adolescents, while Section
\ref{sec:discussion} contains a summary discussion.

\section{Background}\label{sec:background}
The observed data are denoted by $\bx=(x_1,x_2,\cdots,x_m)^T$, an
$m$-dimensional random vector, and the true underlying signals by
$\bs=(s_1,s_2,\cdots,s_n)^T$, an $n$-dimensional transform of
$\bx$. The problem is to determine a matrix $\bW$ so that
\begin{equation}\label{equ:problem}
\bs=\bW \bx.
\end{equation}
A distinguishing feature of ICA components is that the elements of
$\bs$ are assumed to be independent in a generative factor analytic model, instead of
focusing on data-level uncorrelatedness.  The ICA problem can be
formulated using the following generative model for the data:
\begin{equation}\label{equ:model}
\bx= \bA\bs,
\end{equation}
where $\bx$ is the observed $m$-dimensional vector, $\bs$ is the
$n$-dimensional (latent) random vector whose components are assumed
mutually independent, and $\bA$ is a constant $m\times n$ matrix to be
estimated. If it is further assumed that the dimensions of $\bx$ and
$\bs$ are equal, i.e., $ m=n$, the estimate of $\bW$ in
\eqref{equ:problem} is then obtained as the inverse of the estimate
for matrix $\bA$.

Because of the assumption that $\bA$ is invertible, a first stage
dimension reduction is required. Thus the model can be more accurately
stated as 
$\bK\bx=\bA\bs, $ where $\bK$ is a
dimension reduction matrix of size $n\times m$.  Normally, this is
reached via a singular value decomposition of the demeaned data. Henceforth, we will assume
that the data vectors are $n$ dimensional. Also note
that the data are typically de-meaned prior to analysis.

\subsection{FastICA}\label{sec:fastICA}
The previously mentioned dimension reduced data input into an ICA
algorithm are mean zero and uncorrelated, and thus Gaussian
distributional assumptions provide little further insight to linear
reorganizations. This, along with the idea that linearly mixed iid
noise tends to appear Gaussian, has lead researchers to design algorithms that
search for an optimized
$\bW$ that maximizes non-Gaussianity. Such non-Gaussianity of the
independent components is necessary for the identifiability of the
model shown in Equation \eqref{equ:model}
\citep{comon1994independent}. Two common measures of non-Gaussianity
used in ICA are kurtosis \citep{oja2001independent} and
negentropy \citep{pillai2002probability}.

\cite{hyvarinen1999fast} proposed a fast fixed-point algorithm known
as FastICA. Following Hyvarinen's treatment of ICA, mutual information
\citep{oja2001independent}, a natural measure of the dependence
between random variables, can be written as
\[
 J(\by)-\sum\limits_{i} J(y_i),
\]
when the variables $y_i$, $ i=1,\ldots,n$, are uncorrelated. Here
$\by=(y_1,\cdots,y_n)^T$ and $J$ is negentropy, a measure of
non-Gaussianity \citep{comon1994independent}.  Thus, $W$ is determined so that
the mutual information of the independent components, $s_i$, is
minimized. As mentioned by \cite{hyvarinen1999fast}, this is approximately
equivalent to finding directions in which the negentropy is maximized.

For any random variable, $y_i$, with zero mean and unit variance, an approximation
to the negentropy is proportional to $[E\{G(y_i)\}-E\{G(Z)\}]^2$ for appropriately
chosen $G$ where $Z$ is a standard normal. As an example, if $G(y) = y^4$, one results
in ICA methods that approximate minimum kurtosis algorithms.
With this approximation, it can be shown that the problem is equivalent to finding
$
\mbox{argmax}_{w_i} \sum_{i=1}^n J_G(\bw_i)\
$
under the constraint that $E\{(\bw_k^T\bx)(\bw_j^T\bx)\}$ is $0$ with $j\neq k$ and $1$ when $j=k$. Here $J_G$ is given by $J_G(\bw):=[E\{G(\bw^T\bx)\}-E\{G(Z)\}]^2$. 
The FastICA algorithm that we employ uses fixed-point iterations for maximization.

\subsection{Group ICA}\label{sec:groupica}
The extension of ICA to group inferences provides common independent
components across subjects, which allows identification of putative
common brain networks for the whole group.  As in most fMRI studies,
spatial independence is assumed in our group ICA model, since it is
well-suited to the sparse distributed nature of the spatial pattern
for most cognitive activation paradigms \citep{mckeown1997analysis}.
However, instead of the entire brain, the spatial independence
assumption in H-gICA is made for voxels within a single hemisphere.
To use the information of brain functional homotopy, the fMRI data are
registered to symmetric templates and thus the number of voxels are
equal in the left and right hemispheres.  Similarly to in
\cite{calhoun2001method}, a dimension reduction using PCA is applied
to each hemisphere of each subject. Another PCA step is then performed
on the concatenated data matrix, which is not technically necessary
\citep{eloyan2011likelihood} though is the current standard practice
for group ICA.


\section{Methods}\label{sec:methods}
\subsection{Preprocessing}\label{sec:preprocessing}
All group ICA methods for rs-fMRI require brains to be registered
via a deformation algorithm to a canonical brain image, referred to
as a template. This allows the assumption of spatially common
brain networks elaborated on in Section \ref{sec:groupica}. While
our approach is applicable to any standard preprocessing, we
do assume that the template is bilaterally symmetric along the
mid-sagittal plane, contrary to all human neuroanatomy. The
creation and improvement of such templates is beyond the scope of this work.
However briefly, a simple strategy would require an accurate hemispheric
model, a deformation to force the mid-sagittal plane to be perfectly
planar, them flipping one hemisphere and pasting it together with itself
to form the template. Fortunately, symmetric template brains exists;
we employ one create by the International Consortium for 
Brain Mapping (ICBM) \citep{Fonov2011313,Fonov2009S102}.
An easy solution for preprocessing is to take already processed
data, registered to a non-symmetric template, and apply a
second registration to a symmetric template. As such, the
method can be applied to existing processed data easily. 
Without loss of generality, it is assumed that the mid-sagittal
plane is parallel to one of the image dimensions.

\subsection{Homotopic Group ICA}\label{sec:hgICAmodel}
Homotopic group ICA is the simple idea of having each subject
contribute two fMRI images, one for their left hemisphere
and one for their right, to the gICA approach. The benefit
of the approach is the reduction in data level noise by
(effectively) doubling the number of subjects and halving 
the number of voxels. Furthermore, it directly utilizes the
information that many brain networks are apparently bilaterally
symmetric.

Suppose there are $N$ subjects indexed by $i = 1, 2,\ldots ,N$ with
an fMRI time series comprised of $T$ scans.  Further, suppose that
each subject's fMRI image contains $2 V$ voxels with $V$ in each
hemisphere for each time point. Separate the two hemispheres with two
matrices of size $T\times V$, labeled $\bF_{i,1}$ and $\bF_{i,2}$.
Assume these have column sums zero (they are demeaned over time) and
that one side has been appropriately flipped to geometrically match
the other. Conceptually, let $\bF_i(a, b, c, t)$ be a subject's image
represented as a 4D array.  Here, $a$ indexes the left to right
dimension where $a = 1,\ldots, 2K$, $b$ indexes the anterior/posterior dimension, $c$ indexes
the superior/inferior dimension and $t$ indexes scan.  Suppose that
the mid-sagittal plane is defined as $a = K$.
Then, the 4D array defined as $\bF_i(a, b, c, t)$ for $a < K$ defines
one hemisphere's image series, while $\bF_i(2K - a, b, c, t)$ 
defines the other, oriented in a geometrically similar manner. The
left, $\bF_{i,1}$, and right, $\bF_{i,2}$, hemispheric data matrices are obtained 
from the associated left and right
hemispheric images, having vectorized the spatial dimension.

Assume a principal component dimension reduction has been applied and,
abusing notation, that $\bX_{i,j}$ of size $Q\times V$ denotes the
dimension reduced data matrix of the left ($j=1$) and right ($j=2$)
hemispheres of subject $i$.  The column $v$ of $\bX_{i,j}$ continues
to represent voxel $v$ in hemisphere $j$.  The rows of $\bX_{i,j}$ are
the PCs, which are indexed by $t = 1,2,\ldots,Q$.

$\bX_{i,j}(t, v)$ represents row $t$, column $v$ of $\bX_{i,j}$ with
the same convention applied to other vectors and matrices. Assuming
that a group ICA decomposition implies: 
$
\bX_{i,j}(t,v)=\sum\limits_{q=1}^Q \bA_{i,j}(t,q) \bS(q,v),
$
for all $i=1,2,\ldots,N$ and $j=1,2$. This in turn implies that the
spatio-temporal process $\bX_{i,j}(t,v)$ can be decomposed to a
hemisphere specific time series $\bA_{i,j}(t,q)$ and a
subject-independent spatial maps, $\bS(q,v)$. This is equivalent to
the following group ICA model:
\begin{equation}\label{equ:hmodel}
  \bX=\bM\bS.
\end{equation}
Here $\bX=[\bX_1^T,\bX_2^T]^T$ is the $2NQ\times V$ group data matrix
from left and right hemispheres, where $\bX_j :=
[\bX_{1,j}^T,\bX_{2,j}^T,\cdots,\bX_{N,j}^T]^T$, $j=1,2$, which formed
by concatenating $N$ subjects' data in the temporal domain. $\bS$ is
$Q\times V$ matrix containing $Q$ statistically independent spatial
maps in its rows.  $\bM=[\bM_1^T,\bM_2^T]^T$ is the $2NQ\times Q$
group mixing matrix, where
$\bM_j=[\bA_{1,j}^T,\bA_{2,j}^T,\cdots,\bA_{N,j}^T]^T$ is the
$NQ\times Q$ submatrix corresponding to hemisphere $j$ concatenating
the mixing matrix of the $N$ subjects. In the context of fMRI, the
$\bS(q, \cdot)$, $q=1,2,\ldots, Q$ are spatial maps that are the
putative brain networks to be estimated. The mixing matrices
are assumed non singular, and thus we can define
$\bW_{i,j}=\bA_{i,j}^{-1}$.

In the H-gICA model, the independent components are assumed to be
common across subjects and hemispheres, while how they mix to produce
the signal can differ among both subjects and hemispheres.  Of course,
the true mixing matrices are not observed and the FastICA algorithm is
used to obtain the estimates. 

\subsection{Connection with gICA}\label{sec:connection}
In this section H-gICA is linked to the most commonly used group ICA
approach \citep{calhoun2001method}. Let $\tilde{\bX}=[\bX_1,\bX_2]$ be
the $NQ\times 2V$  dimension reduced group data matrix, where $\bX_j = [\bX_{1,j}^T, \bX_{2,j}^T,\cdots,\bX_{N,j}^T]^T$  is same as in Section \ref{sec:hgICAmodel}. Here, we are simply pasting the   
hemispheric data back together.  The standard gICA approach decomposes
$\tilde{\bX}$ as
$
\tilde{\bX}=\tilde{\bM} \tilde{\bS}.
$
Here, $\tilde{\bS}$ is a $Q\times 2V$ matrix containing
$Q$ spatial maps in its rows. And
$\tilde{\bM}=[\tilde{\bA_1}^T,\tilde{\bA_2}^T,\cdots,\tilde{\bA_N}^T]^T$
is the $NQ\times Q$ group mixing matrix, where $\tilde{\bA_i}$ is the
$Q\times Q$ submatrix corresponding to subject $i$. Under the
assumption 
that $\tilde{\bA_i}$, are of full rank,
defines $\tilde{\bW_i}=\tilde{\bA_i}^{-1}$, which can also be
estimated by the FastICA algorithm.  The following theorem, shows that
if the actual sources are truly homotopic and noise free, H-gICA and
gICA provide the same result when using the estimation method of
FastICA. The proof of the theorem is given to the Appendix.

\begin{theorem}\label{thm:same}
  Suppose the actual sources are truly homotopic and noise free. The
  number of estimated ICs is $Q\ll V$. Denote the FastICA estimate
  of $\bS$ to be $\hat{\bS}$ and the estimate of $\tilde{\bS}$ to be
  $\hat{\tilde{\bS}}$. Then for $j\in \{1,2\}$, we have
$
\hat{\tilde{\bS}}=[\hat{\bS},\hat{\bS}].
$
\end{theorem}

Theorem \ref{thm:same} shows that when there is no noise, H-gICA and
GIFT are the same for the homotopic signals. However, of course, noise
exists in most real data sets. Section \ref{sec:simulation} deals with
H-gICA's ability to improve locating underlying sources when noise is
added to the data.

\subsection{Measures of Functional Homotopy}\label{sec:homotopy}
In rs-fMRI, ICA output can be used to obtain measures of network
connectivity, which are of intrinsic interest in studying neural
function \citep{Gao24022012}.  \cite{joel2011relationship} point out
that the functional connectivity between any two brain regions may be
due to within network connectivity (WNC) and between network
connectivity (BNC).  They emphasize the importance of interpreting
such connectivity, ostensibly measured by the correlation and variance
of the temporal mixing matrices.

Similarly to \cite{joel2011relationship} the following defines  a measure of subject- and network-specific functional homotopy for the $k^{th}$  ICA based network of subject $i$:
\begin{equation}\label{equ:homotopy}
H_i(k)=\text{Cor}(\bA_{i,1}^{(k)},\bA_{i,2}^{(k)}),
\end{equation}
where $\bA_{i,j}$ $(i=1,2,\ldots,N, j=1,2)$ 
is the mixing matrix of the $i^{th}$ subject corresponding to hemisphere $j$ and $\bA_{i,j}^{(k)}$ is a vector of the time course modulating spatial map $k$.  Note, typically this requires a back transformation from PCA space.
Here $H_i(k)$ measures the spontaneous activity of the $k^{th}$ network between left and right hemispheres. The estimation of $H_i(k)$ in H-gICA is given by  replacing $\bA_{i,1}$ and $\bA_{i,2}$ with their estimated values in \eqref{equ:homotopy}. Similarly the group level functional homotopy for the $k^{th}$ ICA based network can be define as:
\begin{equation}\label{equ:homotopy2}
H(k)=\text{Cor}(\bM_{1}^{(k)},\bM_{2}^{(k)}),
\end{equation}
where $\bM_j$ is defined in \eqref{equ:hmodel} as the submatrix corresponding to hemisphere $j$ concatenating the mixing matrix of the $N$ subjects and $\bM_j^{(k)}$ is the $t^{th}$ element of the time course modulating spatial map $k$.

\section{Simulation Results}\label{sec:simulation}
\subsection{A Simple Example}
The following simple example is given to initially demonstrate the effectiveness of H-gICA. Suppose the number of subject is $N=3$ and the number of underlying sources is $Q=3$. For each subject $i$, the $T\times 2V$ data matrix $\tilde{\bX}_{i,\cdot} = [\bX_{i,1},\bX_{i,2}]$ is from the model $\tilde{\bX}_{i,\cdot}=\bA_i \tilde{\bS}$ with $T=3$ and $2V=10^2$. We further assume that
\[
\bA_1 = \left( {\begin{array}{*{20}c}
   {-1} & {-5} & {2}  \\
   {5} & {-3} & {2}  \\
   {5} & {3} & {-5}  \\
\end{array}} \right),
\bA_2 = \left( {\begin{array}{*{20}c}
   {-1} & {-1} & {2}  \\
   {-1} & {-2} & {-3}  \\
   {-4} & {0} & {5}  \\
\end{array}} \right),
\bA_3 = \left( {\begin{array}{*{20}c}
   {-3} & {3} & {-5}  \\
   {5} & {-1} & {-1}  \\
   {3} & {-2} & {-1}  \\
\end{array}} \right)
\]
(The elements in $\bA_1$, $\bA_2$ and $\bA_3$ are randomly picked from $Unif(-10,10)$.)

\begin{figure}[htbp]
\begin{center}
\includegraphics[scale=0.25]{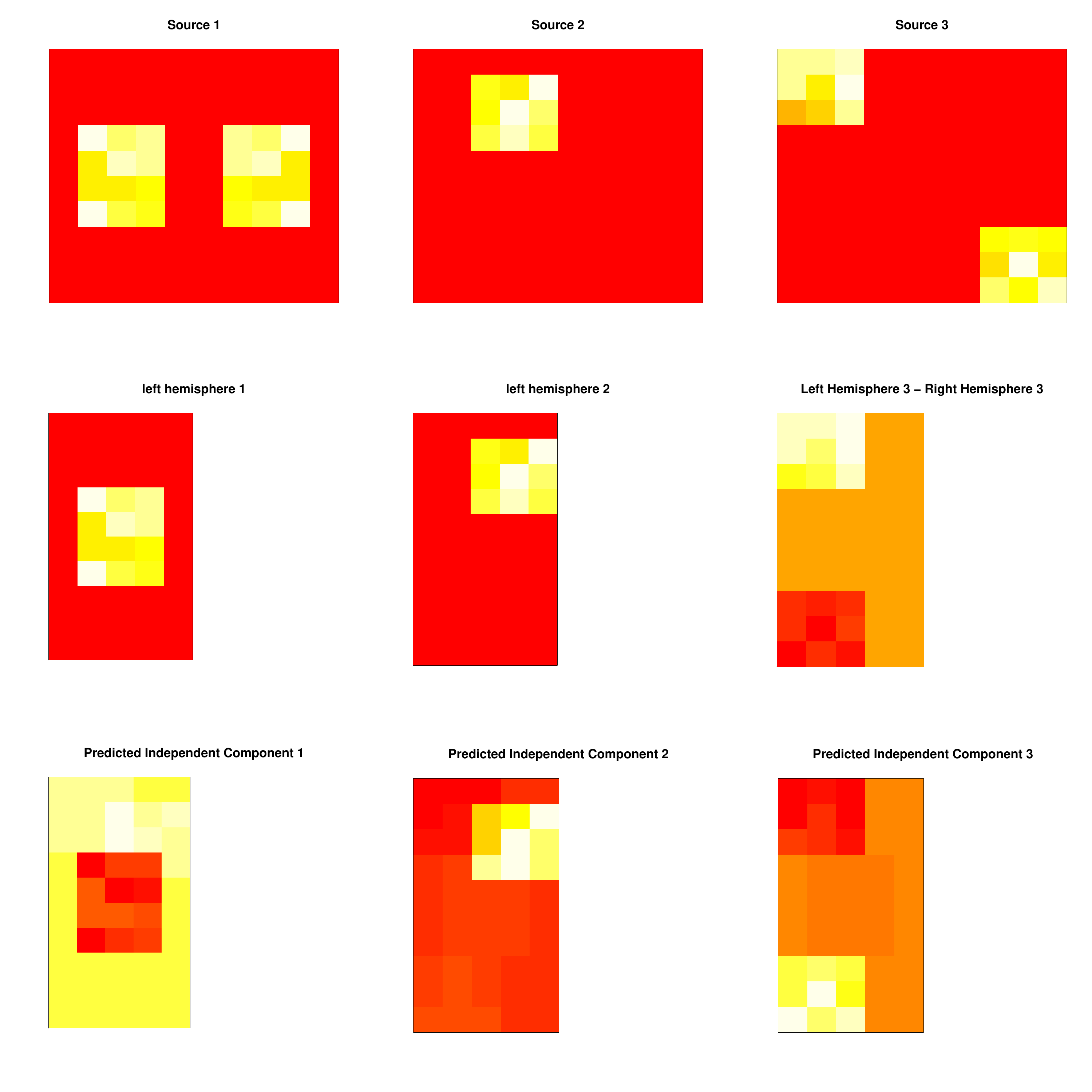}
\caption{Result of the first example. The top row shows the true sources in three different types:  perfectly symmetric;  only present in one hemisphere;  differing in the two hemispheres. The small blocks in these plots are activated voxels, where the values come from uniform distribution. The first two elements of the second row are the true sources in the left hemisphere and the last element is left minus right. The bottom row consists of the independent components generated by H-gICA.}
\label{Fig:toy}
\end{center}
\end{figure}

As shown in Figure \ref{Fig:toy},  three different source matrices ($\tilde{\bS}$) are generated: one is symmetric; another is only in one hemisphere; and the third has two asymmetric blocks of  activated voxels. 
These results are shown in the bottom row of Figure \ref{Fig:toy}, highlighted by H-gICA's separation of the three source matrices and its meaningful prediction for all  of them.

\subsection{2D Simulation}\label{sec:2d_simu}
In order to illustrate the performance of the proposed method,  simulation studies with $10,000$ voxels in 2D spaces were conducted. 
Both the homotopic and non-homotopic settings are used in the study.
 The results are compared with the commonly used group
ICA algorithm without consideration of the left and right hemispheres.

Similar to above,   the number of subjects is $N=3$ and the number of underlying
sources is $Q=3$. The data are generated by the ICA model $\tilde{\bX}_{i,\cdot}=\bA_i \tilde{\bS}$  with $T=3$ and $2V=100^2$.
  $\bA_i$ has the same value as in the toy example for $i=1,2,3$ is also assumed.

\subsubsection{Case I: Perfect Homotopy}  
\begin{figure}[htbp]
\begin{center}
\includegraphics[scale=0.5]{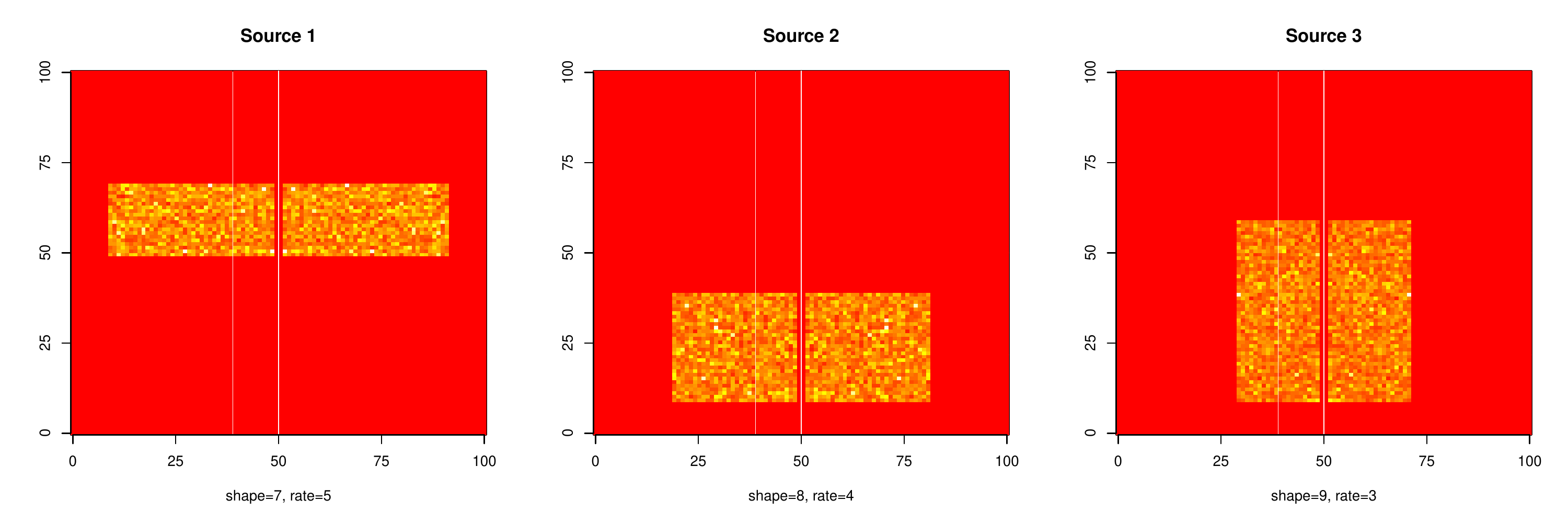}
\caption{True Sources in the Perfect Homotopy. The three sources are generated by  Gamma distributions with different parameters.}
\label{Fig:sympc}
\end{center}
\end{figure}

Assume all  true sources are perfectly homotopic. For each source, two blocks of voxels are symmetrically activated. In each loop of the simulation,  values of these activated voxels are assigned by Gamma distributions. The heatplots of the sources are shown in Figure \ref{Fig:sympc}. 
The data are generated by the three sources in Figure \ref{Fig:sympc} and the mixing matrices $\bA_1$, $\bA_2$, and $\bA_3$. Gaussian noise is then added to the data. The estimated components are standardized and subsequently compared  with the corresponding standardized true sources. Next, the mean difference at each voxel is calculated. The simulation contained 300 iterations in each run.  Homotopic gICA results are compared with the commonly used group ICA algorithm in Table \ref{tab:sym}, where  the noise is set to be mean zero and with a standard deviation equal to $5$.
\begin{table}[htdp]
\begin{center}
\caption{The two rows compare the mean of the voxel mean difference of H-gICA and ordinary gICA in the symmetric setting. }
\vspace{0.2 in}
\begin{tabular}{|c|c|c|c|}
\hline
&{\ttfamily Source 1}&{\ttfamily Source 2}&{\ttfamily Source 3}\\
\hline
H-gICA & 0.414 ($\sim 0.001$) & 0.412 ($\sim 0.001$)& 0.423 ($\sim 0.002$) \\
gICA & 0.507 ($\sim 0.001$)& 0.496 ($\sim 0.001$) & 0.433 ($\sim 0.001$)\\
\hline
\end{tabular}
\label{tab:sym}
\end{center}
\end{table}%

\begin{figure}[htbp]
\begin{center}
\includegraphics[scale=0.6]{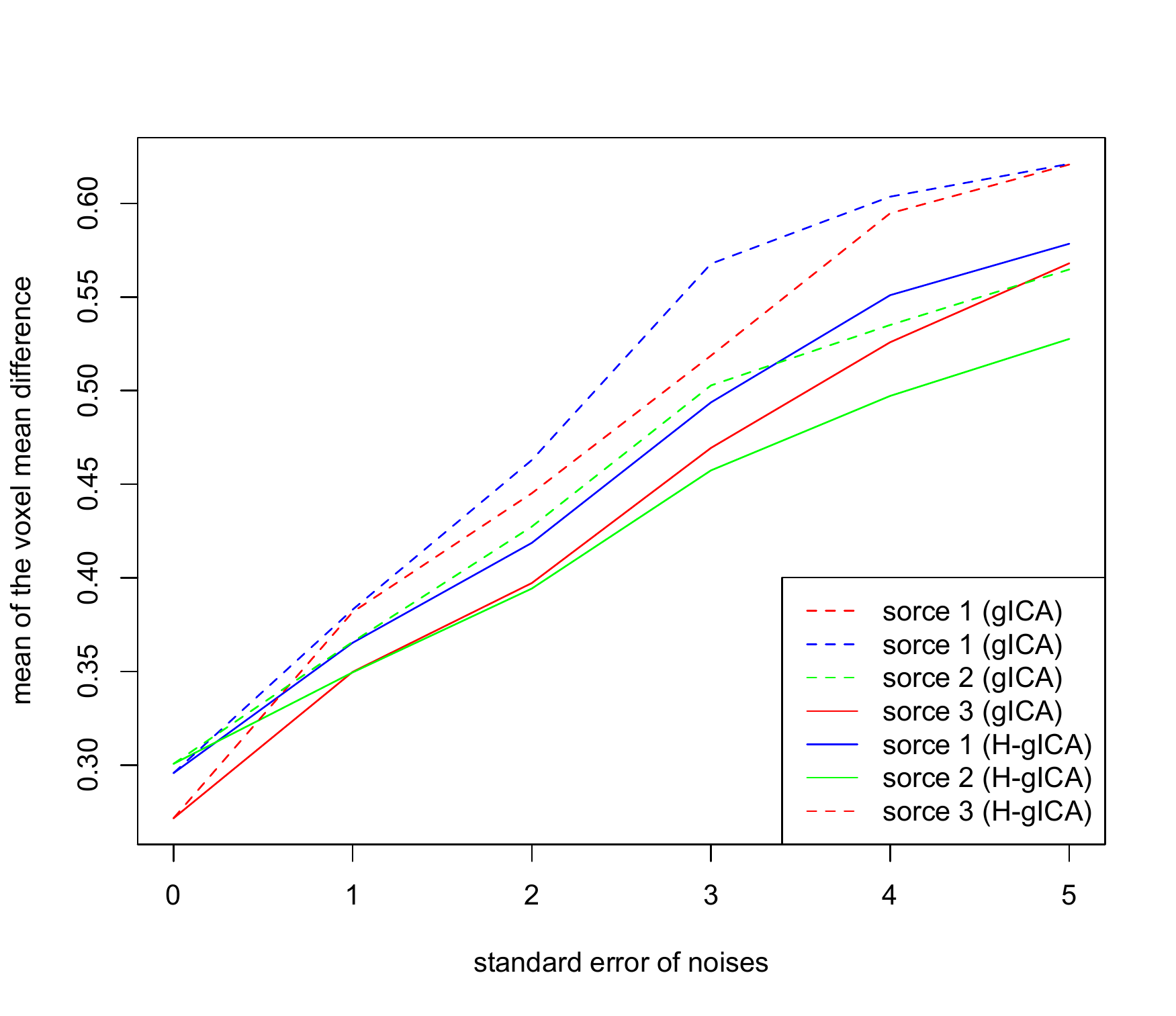}
\caption{The mean of the voxel mean difference increases with the standard deviation of the noise. The two methods are the same under noise-free settings and when the noise exists, H-gICA works better than ordinary gICA.}
\label{Fig:symerr}
\end{center}
\end{figure}

As seen in Table \ref{tab:sym}, in the setting of symmetric sources, the mean errors of H-gICA are smaller than that of ordinary gICA for all the three sources. Figure \ref{Fig:symerr} shows the mean of the voxel mean difference with different settings for the noise. Again, H-gICA works consistently  better when  noise exists and is the same as ordinary gICA in noise-free settings, which was been proven in Theorem \ref{thm:same}.

\begin{figure}[htbp]
\begin{center}
\includegraphics[scale=0.1]{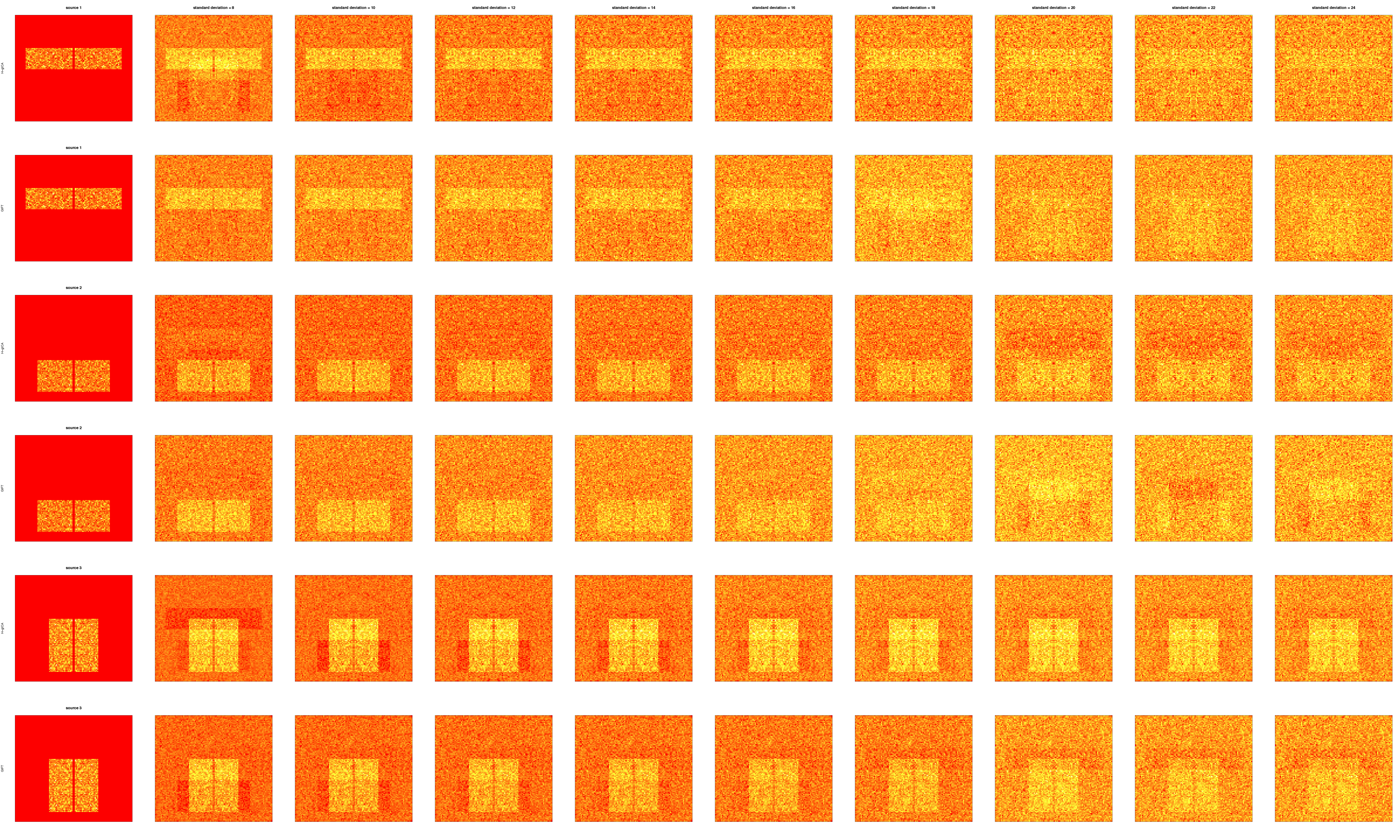}
\caption{The ICs estimated by H-gICA and gICA. The first column shows the true sources. The other columns, from left to right, show the estimated ICs when the noise increases. The $1^{st}$, $3^{rd}$, and $5^{th}$ rows are for H-gICA and the $2_{nd}$, $4^{th}$ and $6^{th}$ are for gICA.}
\label{Fig:ICcompare}
\end{center}
\end{figure}

Figure \ref{Fig:ICcompare} compares the ICs estimated by H-gICA and the ordinary gICA. As we can see, comparing with the ordinary gICA, H-gICA has a consistently better estimation of the ICs when noise exists.

\subsubsection{Case II: Perfect Lateralization}
\begin{figure}[htbp]
\begin{center}
\includegraphics[scale=0.5]{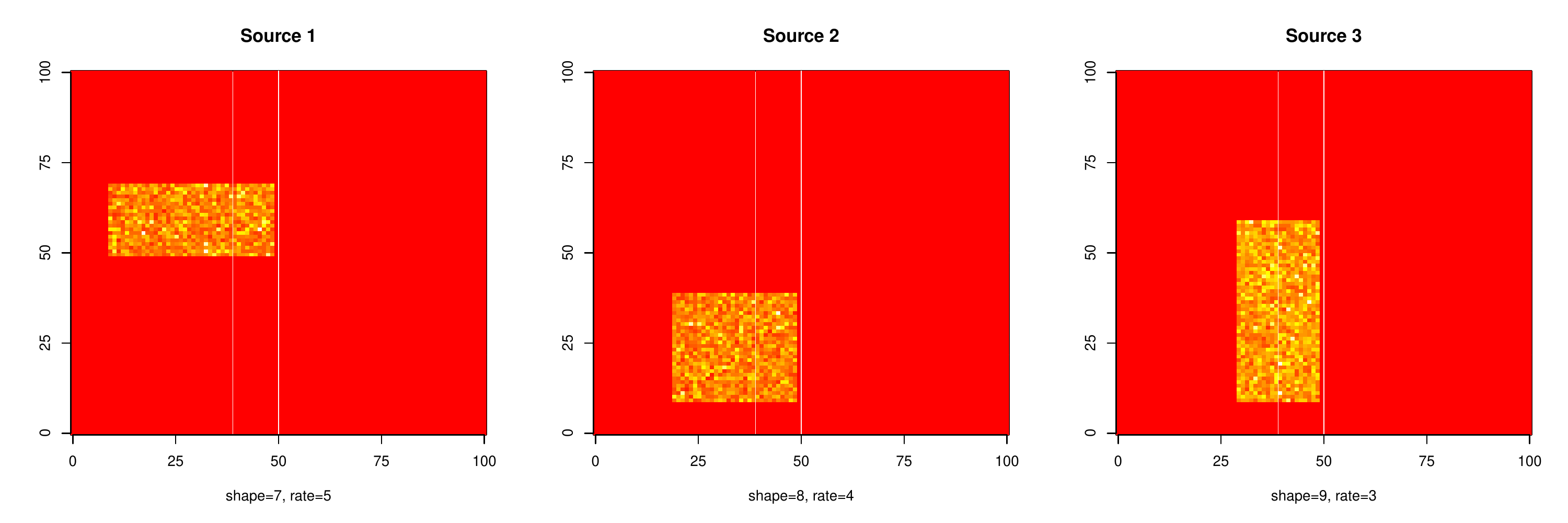}
\caption{True Sources in Non-homotopic Setting. All  three true sources are only in one hemisphere. The values of the activated voxels are generated by Gamma distribution with different parameters.}
\label{Fig:halfpc}
\end{center}
\end{figure}

True sources are now presumed to be only present in one hemisphere (perfect lateralization of the brain network), and without loss of generality,  the left hemisphere is selected.  Similar to Case I, a Gamma distribution was used to generate the value of the activated voxels and, again, 300 iterations were  run. Figure \ref{Fig:halfpc}  shows the heatplots of the three true sources in the first iteration.
Gassian noise was added to the data generated by the three sources and the results are compared with ordinary gICA. Note that, since the independent components provided by H-gICA contain only one hemisphere by design. Thus, the sole applicable comparison is the true sources in the left hemisphere.  Table \ref{tab:half} gives these results, and show that the mean error of H-gICA is less than that of  gICA for  Source 2 while larger than the mean error of  gICA for Source 1 and Source 3. In summary, H-gICA is competitive with normal g-ICA when the sources are lateralized with the caveat that H-gICA does not provide lateralization information on which hemisphere the network resides in. We do note, however, that this information is contained in the temporal mixing matrix, just not in a form easily displayed as an image.
\begin{table}[htdp]
\begin{center}
\caption{Comparison of the H-gICA results versus ordinary gICA results when the true sources are only in one hemisphere. The table provides the mean and standard deviation  of the voxel mean difference with the true sources for the 300 iterations.  }
\vspace{0.2 in}
\begin{tabular}{|c|c|c|c|}
\hline
&{\ttfamily Source 1}&{\ttfamily Source 2}&{\ttfamily Source 3}\\
\hline
H-gICA & 0.508 ($\sim 0.002$) & 0.494 ($\sim 0.001$) & 0.431 ($\sim 0.002$)    \\
gICA & 0.502 ($\sim 0.003$)   & 0.533 ($\sim 0.002$)& 0.312 ($\sim 0.003$)  \\
\hline
\end{tabular}
\label{tab:half}
\end{center}
\end{table}%

\subsubsection{Case III: Mixture of Lateralized and Homotopic Networks}
\begin{figure}[htbp]
\begin{center}
\includegraphics[scale=0.5]{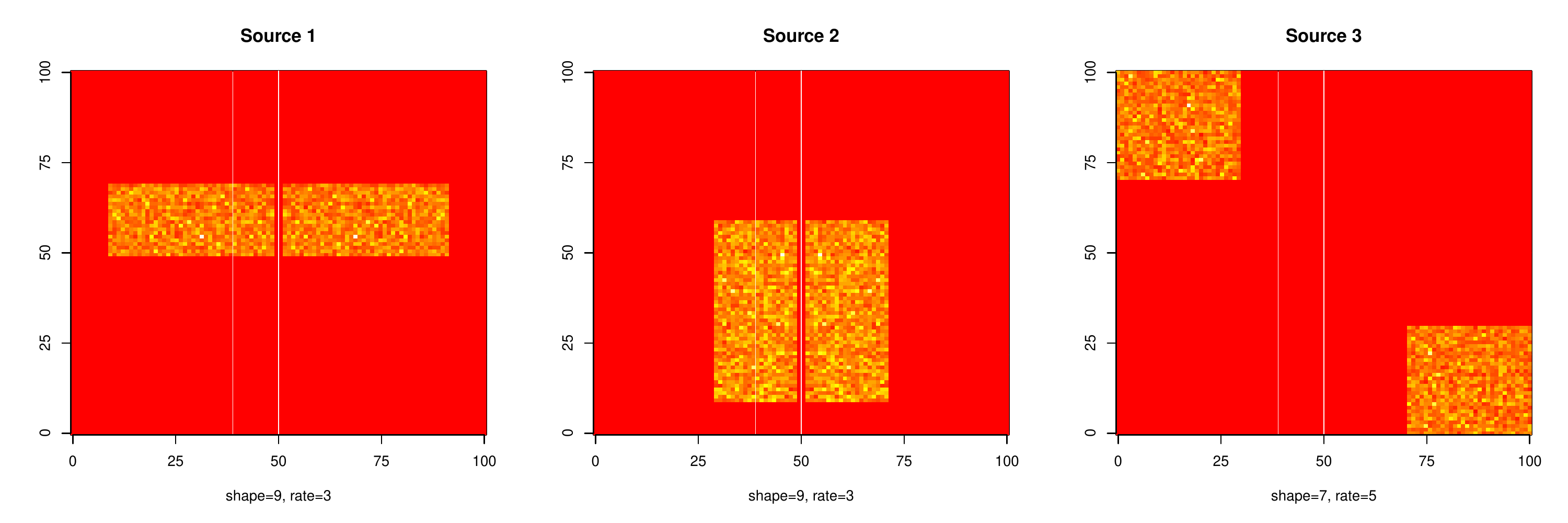}
\caption{True Sources in Mixed Setting. The first two sources are homotopic while the third one varies in the two hemispheres.}
\label{Fig:mixedpc}
\end{center}
\end{figure}

\begin{figure}[htbp]
\begin{center}
\includegraphics[scale=0.1]{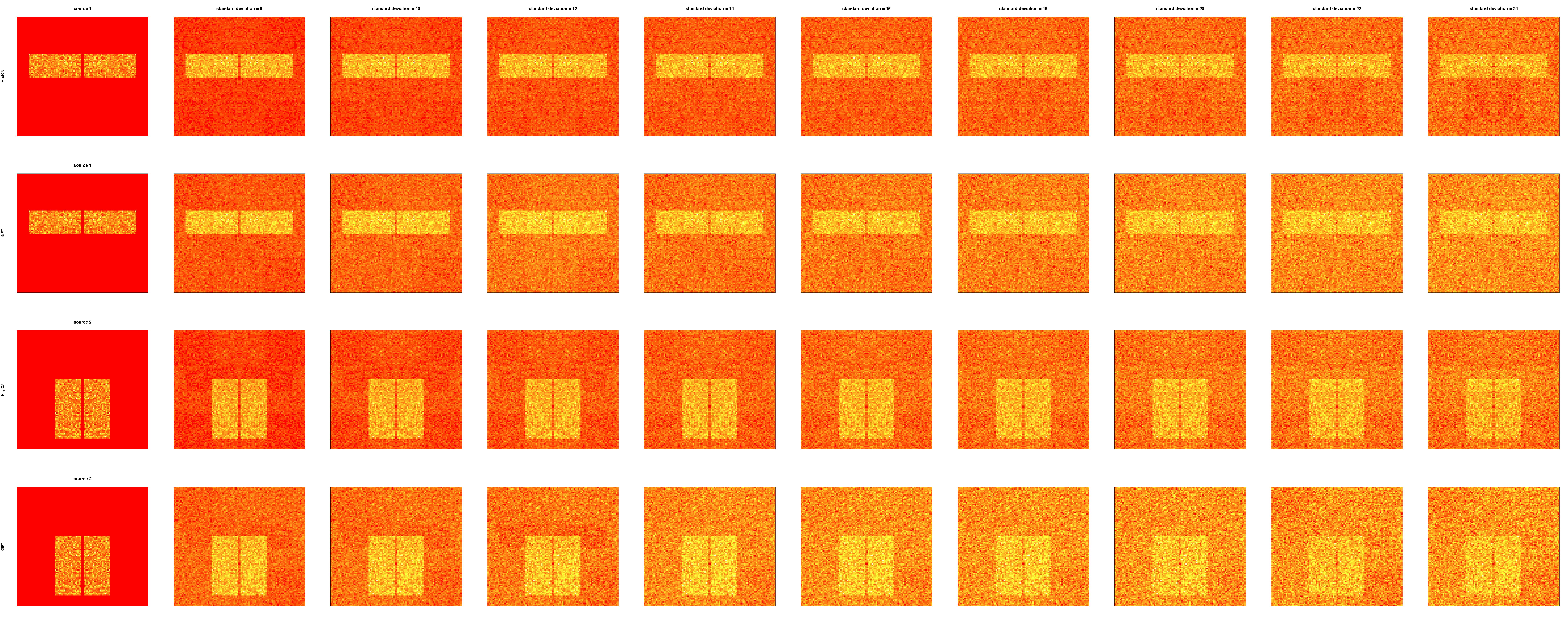}
\caption{The homotopic ICs estimated by H-gICA and ordinary gICA. The first column shows the true sources. The other columns, from left to right, contain the estimated ICs with increasing  noise. The $1^{st}$ and $3^{rd}$ rows are results of H-gICA and  the $2_{nd}$ and $4^{th}$ are of ordinary gICA.}
\label{Fig:ICcompare_mixed}
\end{center}
\end{figure}

For this setting,  a mixture of two different types of sources is introduced. As shown in Figure \ref{Fig:mixedpc}, the first two sources are homotopic while the third source covers different regions in the two hemispheres. The estimated ICs for the two homotopic sources are in Figure \ref{Fig:ICcompare_mixed}, which illustrates that the effectiveness of estimating the homotopic sources is not impacted by adding non-homotopic sources. 

\subsection{Flag Example}
\begin{figure}[htbp]
\begin{center}
\includegraphics[scale=0.5]{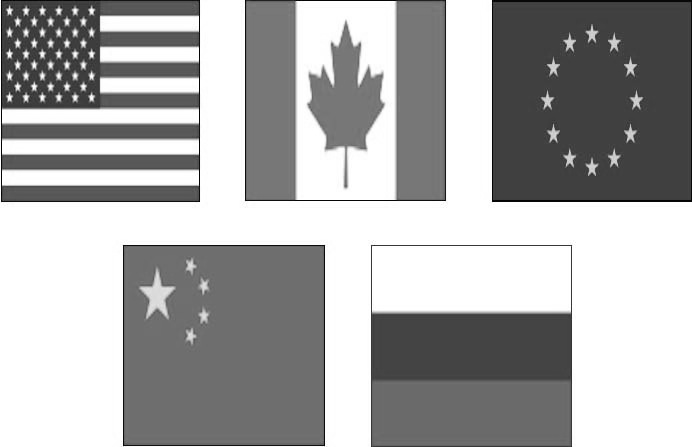}
\caption{Actual sources of the flag example. The original flags images were taken from Wikipedia.}
\label{Fig:flags}
\end{center}
\end{figure}

\begin{figure}[htbp]
\begin{center}
\includegraphics[scale=0.45]{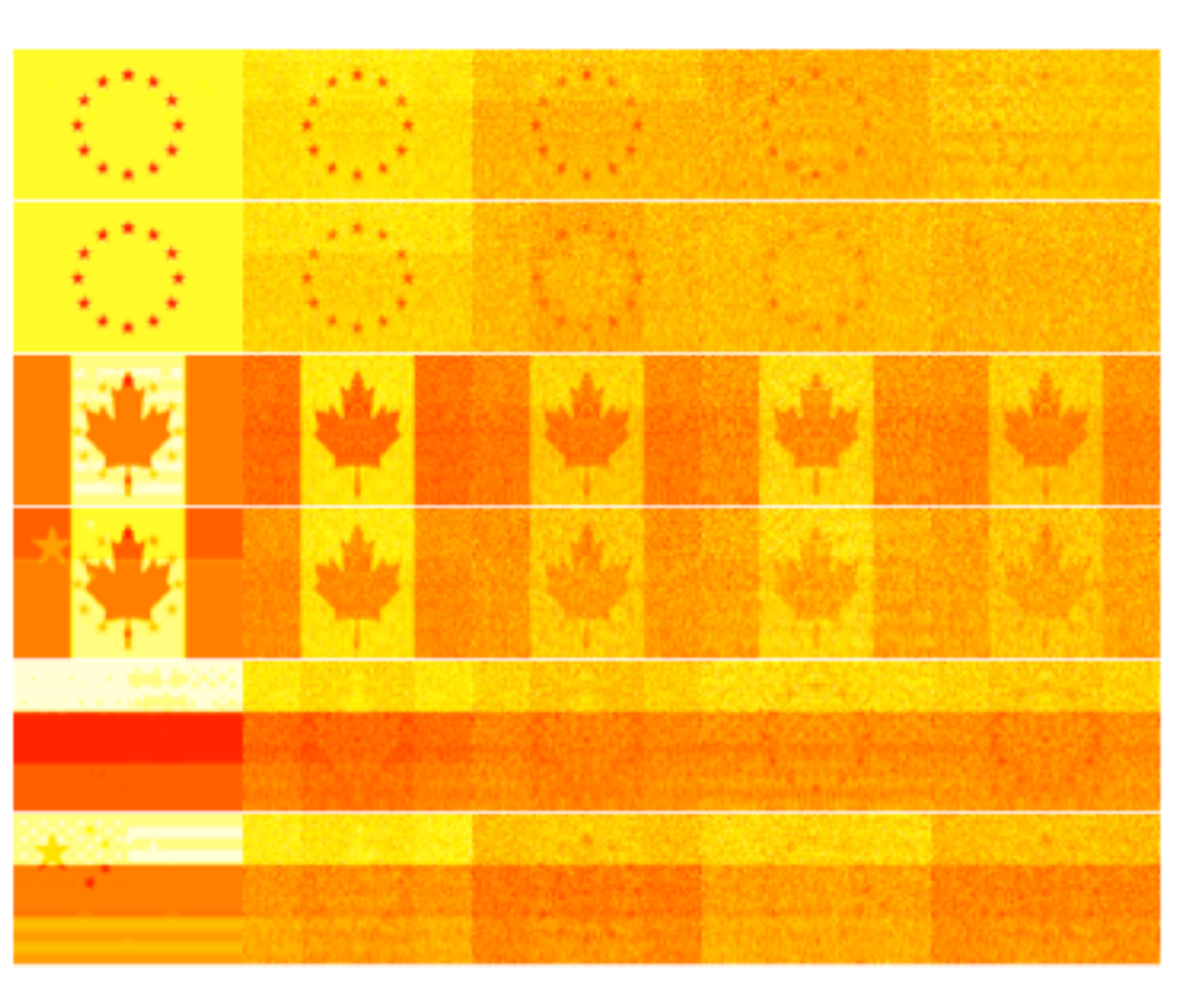}
\caption{The estimated homotopic sources of H-gICA and ordinary gICA. The five columns, from left to right, contain the estimated ICs with increasing noise. The $1^{st}$, $3^{rd}$ and $5^{th}$ rows shows the estimated sources extracted by H-gICA while the $2_{nd}$, $4^{th}$ and $6^{th}$ rows shows the estimated sources extracted by ordinary gICA.}
\label{Fig:flags_results}
\end{center}
\end{figure}

In this example, the actual sources will be the gray-scaled flags of the USA, Canada, the European Union, China and Russia. As shown in Figure \ref{Fig:flags}, three of them are symmetric (Canada, the European Union and Russia) and two are not (USA and China). Similar as in Section \ref{sec:2d_simu}, the data are generated by the ICA model with a fixed mixing matrix, Gaussian noise was then added. The results are shown in Figure \ref{Fig:flags_results}, where the $1^{st}$, $3^{rd}$ and $5^{th}$ rows are the estimated symmetric sources  extracted by H-gICA and the $2_{nd}$, $4^{th}$ and $6^{th}$ rows are the estimated symmetric sources extracted by ordinary gICA. As we can see, for all of the three symmetric sources, H-gICA provides clearer estimation as the noises increased. Moreover, leakage that is apparent in the g-ICA is not apparent in H-gICA.

\section{Application to the ADHD-200 Dataset}\label{sec:application}
Application of the H-gICA method is illustrated using the ADHD-200 dataset, one of the largest and freely available resting state fMRI datasets. The data combines 776 resting-state fMRI and anatomical scans aggregated across eight independent imaging sites,
491 of which were obtained from normally developing individuals and 285 from children and
adolescents with ADHD (ages: 7-21 years old). We view this analysis as largely a proof of principle in applying the method and defer thorough investigations of ADHD for later work.

\begin{figure}[htbp]
\begin{center}
\includegraphics[scale=0.45]{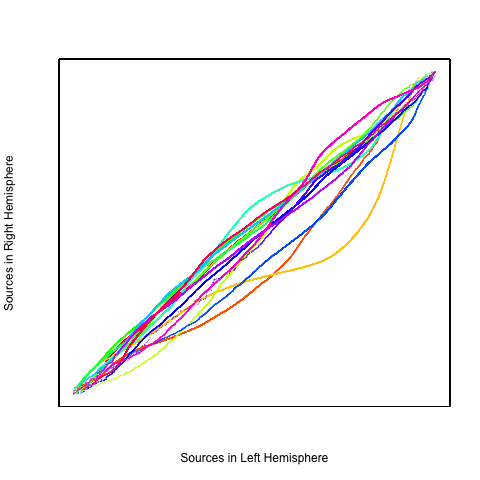}
\caption{QQ plot of the 15 sources extracted by gICA.}
\label{Fig:qq}
\end{center}
\end{figure}

\begin{figure}[htbp]
\begin{center}
\includegraphics[scale=0.35]{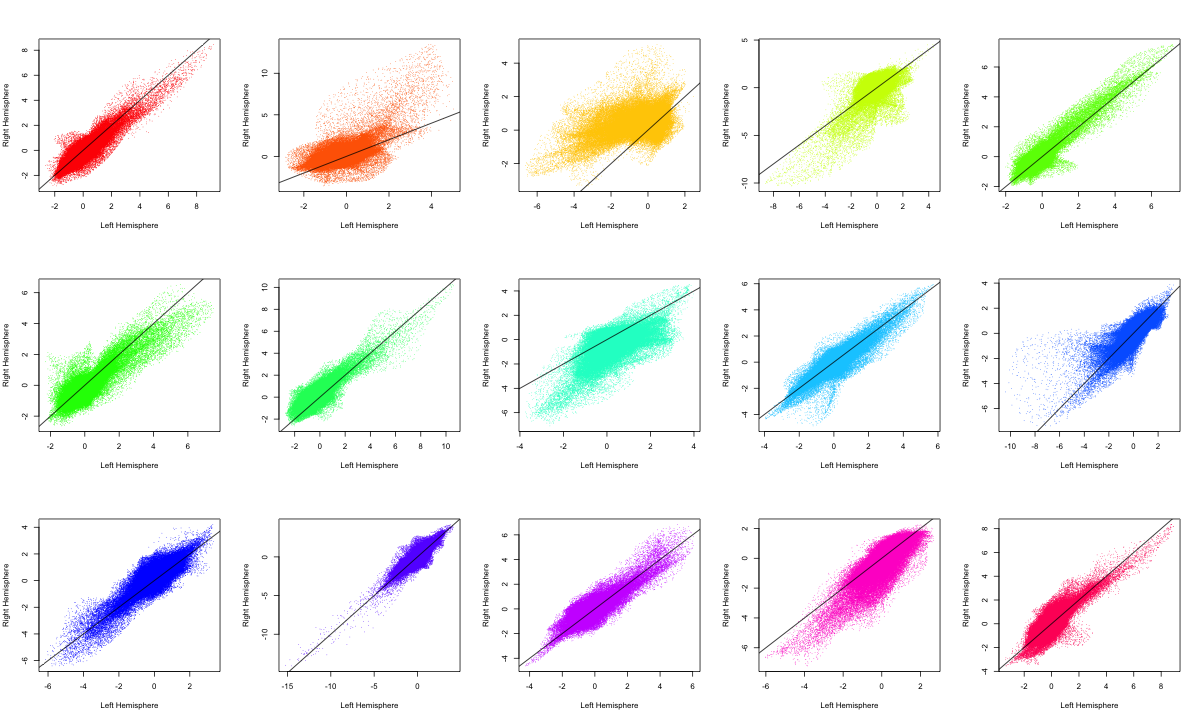}
\caption{The scatter plot of the 15 sources.}
\label{Fig:sources15}
\end{center}
\end{figure}

This particular analysis focused on 20 subjects picked from the ADHD-200 data set.
Data were processed via the NITRC 1,000 Functional Connectome processing scripts \citep{mennes2012making}. 
In summary, images were slice-time corrected, deobliqued, skull stripped, smoothed and registered to a 3 mm3 MNI template. 
The data were then registered to ICBM 2009a nonlinear symmetric
templates generated by the McConnell Brain Imaging Centre
\citep{fonov2009unbiased,fonov2011unbiased}. Each fMRI scan contains
$99\times 117\times95=1,100,385$ voxels measured at $176$ time
points. Figure \ref{Fig:qq} and Figure \ref{Fig:sources15} are the QQ
plot and scatter plots of the estimated sources extracted by ordinary
gICA in the left and right hemispheres. Most of the estimated sources
are close to the $45^o$ line, which suggests that the marginal
distributions of left and right hemispheres are similar.  H-gICA can
benefit from this apparent homotopy.

\begin{figure}[htbp]
\begin{center}
	\subfigure[]{
		\label{fig:visual}
		\includegraphics[width=0.47\textwidth]{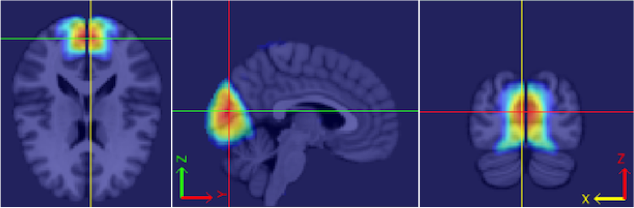}
	}
	\subfigure[]{
		\label{fig:visual_gift}
		\includegraphics[width=0.47\textwidth]{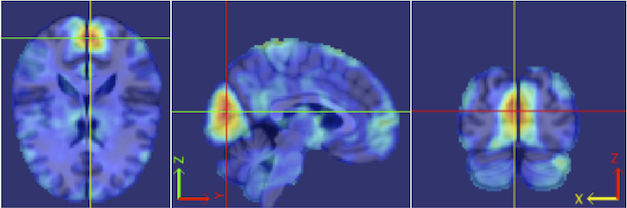}
	}\\
	\subfigure[]{
		\label{fig:default}
		\includegraphics[width=0.47\textwidth]{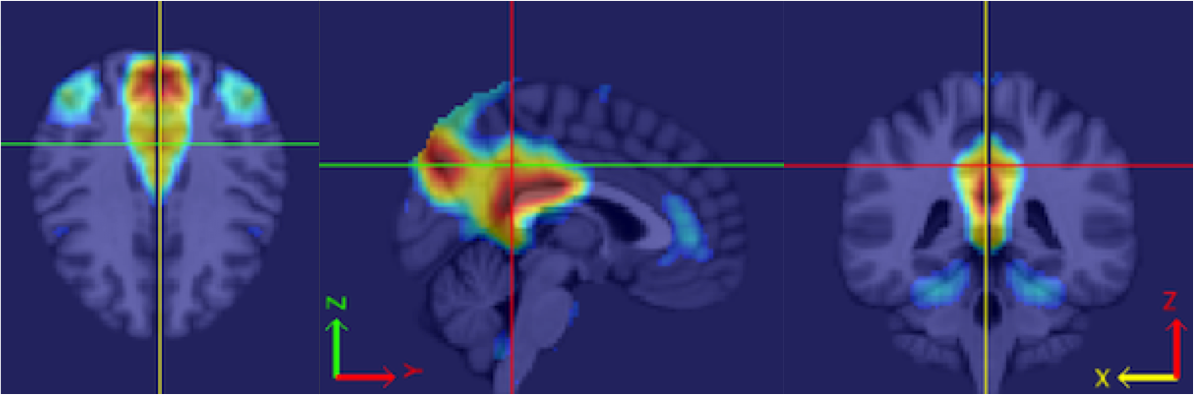}
	}
	\subfigure[]{
		\label{fig:default_gift}
		\includegraphics[width=0.47\textwidth]{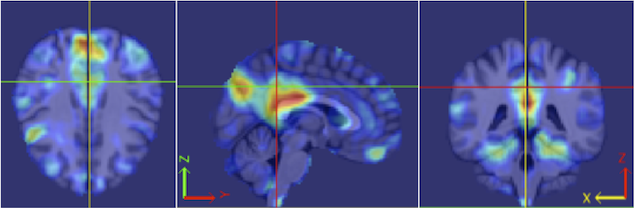}
	}\\
	\subfigure[]{
		\label{fig:auditory}
		\includegraphics[width=0.47\textwidth]{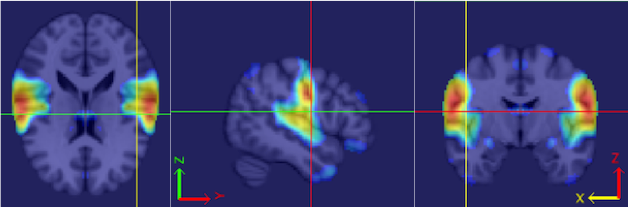}
	}
	\subfigure[]{
		\label{fig:auditory_gift}
		\includegraphics[width=0.47\textwidth]{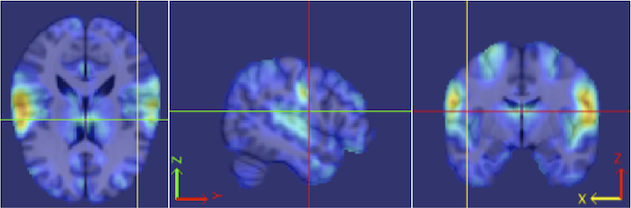}
	}\\
	\subfigure[]{
		\label{fig:motor}
		\includegraphics[width=0.47\textwidth]{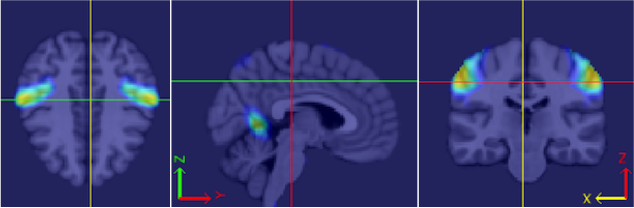}
	}
	\subfigure[]{
		\label{fig:motor_gift}
		\includegraphics[width=0.47\textwidth]{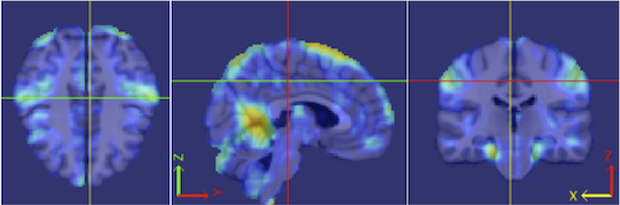}
	}\\
\end{center}
\caption{Comparison of ICs obtain from H-gICA (left column) and ordinary gICA (right column).}\label{fig:components}
\end{figure}

Our procedure is as follows. Each fMRI scan was separated into left
and right hemispheres. Thus, each hemisphere contained $49\times
117\times95=544,635$ voxels. Similar to standard group ICA
\citep{calhoun2001method}, a dimension reduction using PCA was applied
to each hemisphere of each subject. 15 PCs are obtained for each
hemisphere. A group data matrix was generated by concatenating the
reduced data of both hemispheres of the 20 subjects in the temporal
domain. Thus, the aggregated matrix has dimension $2NT\times V$, where
$N=20$, $T=15$, and $V=544,635$.  Our algorithm of homotopic group ICA
is then applied on this matrix. Fifteen estimated independent
components are postulated by H-gICA. As shown in Figure
\ref{fig:components}, out of the 15 components, several brain networks
were found including: the visual network \ref{fig:visual}, the default
mode network \ref{fig:default}, the auditory network
\ref{fig:auditory}, and the motor network \ref{fig:motor}.  Compared
with the ICs obtained from ordinary gICA, shown in
\ref{fig:visual_gift}, \ref{fig:default_gift}, \ref{fig:auditory_gift}
and \ref{fig:motor_gift}, H-gICA improves the estimation of all of
these sources by yielding substantially more clearly delineated
networks.

\begin{figure}[htbp]
\begin{center}
\includegraphics[width=0.8\textwidth]{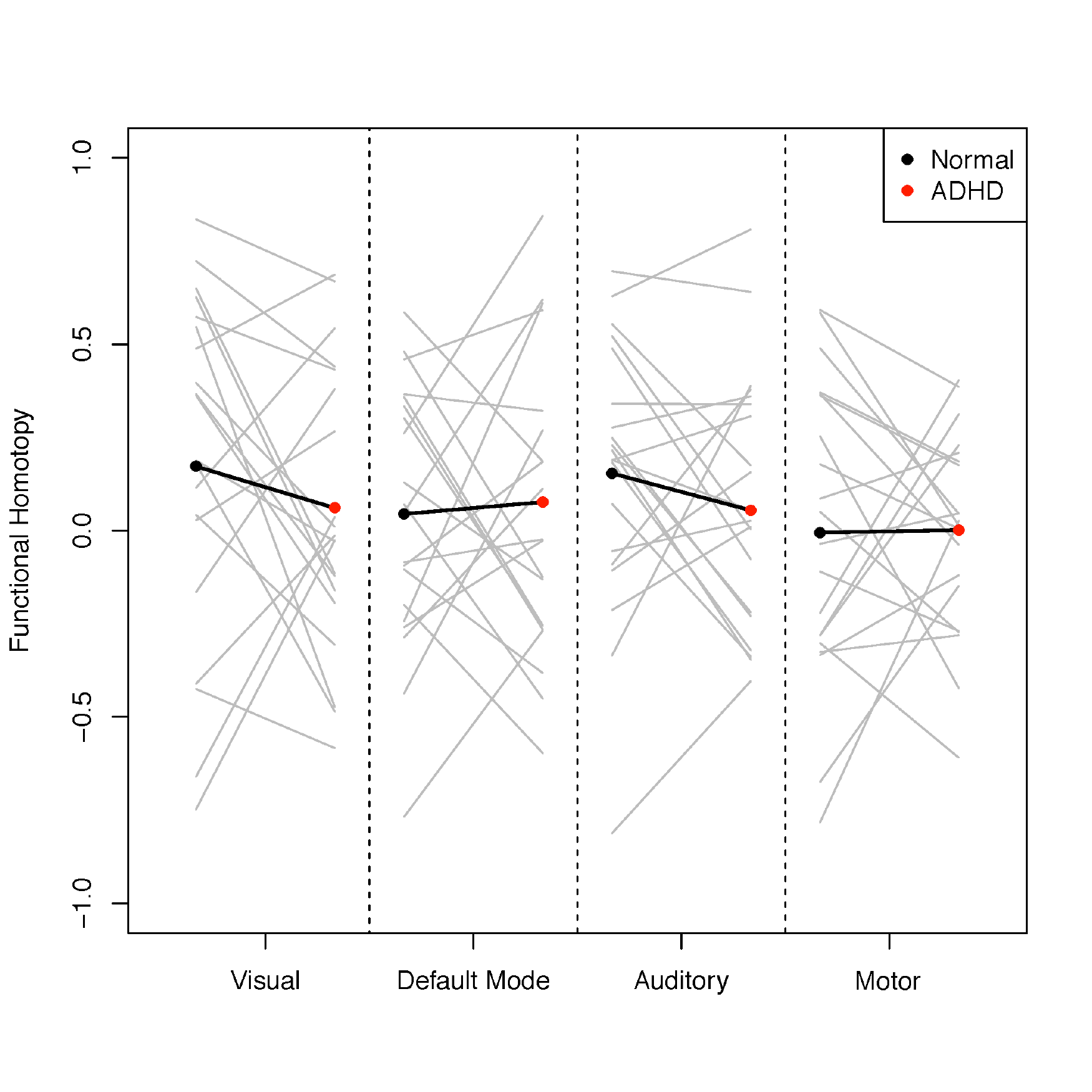}
\caption{Comparison of  functional homotopy of ADHD and normal developed children. Each column represents a network (visual, default mode, auditory and motor). Each pair (subjects and controls) are connected via a grey line. The left end points of the grey lines measure the functional homotopy of the control and the right end points are for the ADHD subjects. The black lines represent the group level functional homotopy for the four networks.}
\label{fig:homotopy}
\end{center}
\end{figure}

The approach of H-gICA also allows us to calculate the brain functional homotopy of each brain network. To compare the  brain functional homotopy of ADHD and typical developed children, we choose 20 ADHD subjects and 20 subject-macthed controls. The subjects and controls were matched in gender and age. Via Equations  \eqref{equ:homotopy} and \eqref{equ:homotopy2}, the estimated  functional homotopy of four networks (visual, default mode, auditory and motor) are shown in Figure \ref{fig:homotopy}. As we can see, the functional homotopy of ADHD children tends to be lower in both visual networks and the auditory network. These represent meaningful leads on the exploration of homotopic network relationships and disease, though we leave a full exploration to later work.

\section{Discussion}\label{sec:discussion}
In this paper we present a new group ICA method called homotopic group ICA (H-gICA). Similar with ordinary group ICA methods, H-gICA can analyze data for multiple subjects
and estimate common underlying IC's across individuals.
By concatenating the fMRI data of the two hemispheres, H-gICA effectively doubles the sample size. It improves the power of finding the underlying brain networks by  rearranging the data structure and utilizing the known information of synchrony between bi-laterally symmetrically opposing inter-hemispheric regions. 
Both the simulation study and the application on ADHD 200 data show that H-gICA is preferable to ordinary gICA when the data are homotopic, increasingly so as
noise increases. Moreover, H-gICA remains preferable to gICA at estimating homotopic sources, even in the presence of non-homotopic sources.   Effectiveness was demonstrated by application on the ADHD-200 dataset. Several brain networks were found and clearly represented in smoother, more clearly delineated, contiguous volumes than ordinary gICA. The main networks found included  visual networks, the default mode network, the auditory network, as well as others.
Moreover H-gICA enables certain measurement of the functional homotopy of the underlying functional networks.  This potentially offers the opportunity to analyze the relation of the brain functional homotopy between the left and right  hemisphere of the brain with diseases.

\newpage

\bibliography{mybib}

\begin{thebibliography}{31}
\providecommand{\natexlab}[1]{#1}
\providecommand{\url}[1]{\texttt{#1}}
\expandafter\ifx\csname urlstyle\endcsname\relax
  \providecommand{\doi}[1]{doi: #1}\else
  \providecommand{\doi}{doi: \begingroup \urlstyle{rm}\Url}\fi

\bibitem[Beckmann and Smith(2004)]{beckmann2004probabilistic}
C.F. Beckmann and S.M. Smith.
\newblock Probabilistic independent component analysis for functional magnetic
  resonance imaging.
\newblock \emph{Medical Imaging, IEEE Transactions on}, 23\penalty0
  (2):\penalty0 137--152, 2004.

\bibitem[Beckmann and Smith(2005)]{beckmann2005tensorial}
C.F. Beckmann and S.M. Smith.
\newblock Tensorial extensions of independent component analysis for
  multisubject fmri analysis.
\newblock \emph{Neuroimage}, 25\penalty0 (1):\penalty0 294--311, 2005.

\bibitem[Biswal et~al.(1995)Biswal, Zerrin~Yetkin, Haughton, and
  Hyde]{biswal1995functional}
B.~Biswal, F.~Zerrin~Yetkin, V.M. Haughton, and J.S. Hyde.
\newblock Functional connectivity in the motor cortex of resting human brain
  using echo-planar mri.
\newblock \emph{Magnetic resonance in medicine}, 34\penalty0 (4):\penalty0
  537--541, 1995.

\bibitem[Calhoun et~al.(2001{\natexlab{a}})Calhoun, Adali, McGinty, Pekar,
  Watson, Pearlson, et~al.]{calhoun2001fmri}
VD~Calhoun, T.~Adali, VB~McGinty, JJ~Pekar, TD~Watson, GD~Pearlson, et~al.
\newblock fmri activation in a visual-perception task: network of areas
  detected using the general linear model and independent components analysis.
\newblock \emph{NeuroImage}, 14\penalty0 (5):\penalty0 1080--1088,
  2001{\natexlab{a}}.

\bibitem[Calhoun et~al.(2001{\natexlab{b}})Calhoun, Adali, Pearlson, and
  Pekar]{calhoun2001method}
VD~Calhoun, T.~Adali, GD~Pearlson, and JJ~Pekar.
\newblock A method for making group inferences from functional mri data using
  independent component analysis.
\newblock \emph{Human brain mapping}, 14\penalty0 (3):\penalty0 140--151,
  2001{\natexlab{b}}.

\bibitem[Comon(1994)]{comon1994independent}
P.~Comon.
\newblock Independent component analysis, a new concept?
\newblock \emph{Signal processing}, 36\penalty0 (3):\penalty0 287--314, 1994.

\bibitem[Damoiseaux et~al.(2008)Damoiseaux, Beckmann, Arigita, Barkhof,
  Scheltens, Stam, Smith, and Rombouts]{damoiseaux2008reduced}
JS~Damoiseaux, CF~Beckmann, E.J.S. Arigita, F.~Barkhof, P.~Scheltens, CJ~Stam,
  SM~Smith, and S.~Rombouts.
\newblock Reduced resting-state brain activity in the Òdefault networkÓ in
  normal aging.
\newblock \emph{Cerebral Cortex}, 18\penalty0 (8):\penalty0 1856--1864, 2008.

\bibitem[Eloyan et~al.(2012)Eloyan, Muschelli, Nebel, Liu, Han, Zhao, Barber,
  Joel, Pekar, Mostofsky, et~al.]{eloyan2012automated}
Ani Eloyan, John Muschelli, Mary~Beth Nebel, Han Liu, Fang Han, Tuo Zhao,
  Anita~D Barber, Suresh Joel, James~J Pekar, Stewart~H Mostofsky, et~al.
\newblock Automated diagnoses of attention deficit hyperactive disorder using
  magnetic resonance imaging.
\newblock \emph{Frontiers in Systems Neuroscience}, 6:\penalty0 61, 2012.

\bibitem[Eloyan et~al.(2013)Eloyan, Crainiceanu, and
  Caffo]{eloyan2011likelihood}
Ani Eloyan, Ciprian~M. Crainiceanu, and Brian~S. Caffo.
\newblock Likelihood-based population independent component analysis.
\newblock \emph{Biostatistics}, 2013.

\bibitem[Esposito et~al.(2005)Esposito, Scarabino, Hyvarinen, Himberg,
  Formisano, Comani, Tedeschi, Goebel, Seifritz, Di~Salle,
  et~al.]{esposito2005independent}
F.~Esposito, T.~Scarabino, A.~Hyvarinen, J.~Himberg, E.~Formisano, S.~Comani,
  G.~Tedeschi, R.~Goebel, E.~Seifritz, F.~Di~Salle, et~al.
\newblock Independent component analysis of fmri group studies by
  self-organizing clustering.
\newblock \emph{Neuroimage}, 25\penalty0 (1):\penalty0 193--205, 2005.

\bibitem[Fonov et~al.(2011{\natexlab{a}})Fonov, Evans, Botteron, Almli,
  McKinstry, and Collins]{fonov2011unbiased}
V.~Fonov, A.C. Evans, K.~Botteron, C.R. Almli, R.C. McKinstry, and D.L.
  Collins.
\newblock Unbiased average age-appropriate atlases for pediatric studies.
\newblock \emph{NeuroImage}, 54\penalty0 (1):\penalty0 313, 2011{\natexlab{a}}.

\bibitem[Fonov et~al.(2011{\natexlab{b}})Fonov, Evans, Botteron, Almli,
  McKinstry, and Collins]{Fonov2011313}
Vladimir Fonov, Alan~C Evans, Kelly Botteron, C~Robert Almli, Robert~C
  McKinstry, and D~Louis Collins.
\newblock Unbiased average age-appropriate atlases for pediatric studies.
\newblock \emph{NeuroImage}, 54\penalty0 (1):\penalty0 313, 2011{\natexlab{b}}.

\bibitem[Fonov et~al.(2009{\natexlab{a}})Fonov, Evans, McKinstry, Almli, and
  Collins]{Fonov2009S102}
VS~Fonov, AC~Evans, RC~McKinstry, CR~Almli, and DL~Collins.
\newblock Unbiased nonlinear average age-appropriate brain templates from birth
  to adulthood.
\newblock \emph{NeuroImage}, 47:\penalty0 S102, 2009{\natexlab{a}}.

\bibitem[Fonov et~al.(2009{\natexlab{b}})Fonov, Evans, McKinstry, Almli, and
  Collins]{fonov2009unbiased}
VS~Fonov, AC~Evans, RC~McKinstry, CR~Almli, and DL~Collins.
\newblock Unbiased nonlinear average age-appropriate brain templates from birth
  to adulthood.
\newblock \emph{Neuroimage}, 47:\penalty0 S102, 2009{\natexlab{b}}.

\bibitem[Gao et~al.(2009)Gao, Zhu, Giovanello, Smith, Shen, Gilmore, and
  Lin]{gao2009evidence}
Wei Gao, Hongtu Zhu, Kelly~S Giovanello, J~Keith Smith, Dinggang Shen, John~H
  Gilmore, and Weili Lin.
\newblock Evidence on the emergence of the brain's default network from
  2-week-old to 2-year-old healthy pediatric subjects.
\newblock \emph{Proceedings of the National Academy of Sciences}, 106\penalty0
  (16):\penalty0 6790--6795, 2009.

\bibitem[Gao et~al.(2012)Gao, Gilmore, Shen, Smith, Zhu, and Lin]{Gao24022012}
Wei Gao, John~H Gilmore, Dinggang Shen, Jeffery~Keith Smith, Hongtu Zhu, and
  Weili Lin.
\newblock The synchronization within and interaction between the default and
  dorsal attention networks in early infancy.
\newblock \emph{Cerebral Cortex}, 2012.

\bibitem[Garrity et~al.(2007)Garrity, Pearlson, McKiernan, Lloyd, Kiehl, and
  Calhoun]{garrity2007aberrant}
A.~Garrity, G.~Pearlson, K.~McKiernan, D.~Lloyd, K.~Kiehl, and V.~Calhoun.
\newblock Aberrant Òdefault modeÓ functional connectivity in schizophrenia.
\newblock \emph{American Journal of Psychiatry}, 164\penalty0 (3):\penalty0
  450--457, 2007.

\bibitem[Guo and Pagnoni(2008)]{guo2008unified}
Y.~Guo and G.~Pagnoni.
\newblock A unified framework for group independent component analysis for
  multi-subject fmri data.
\newblock \emph{NeuroImage}, 42\penalty0 (3):\penalty0 1078--1093, 2008.

\bibitem[Gusnard et~al.(2001)Gusnard, Raichle, Raichle,
  et~al.]{gusnard2001searching}
D.A. Gusnard, M.E. Raichle, ME~Raichle, et~al.
\newblock Searching for a baseline: functional imaging and the resting human
  brain.
\newblock \emph{Nature Reviews Neuroscience}, 2\penalty0 (10):\penalty0
  685--694, 2001.

\bibitem[Hyv{\"a}rinen(1999)]{hyvarinen1999fast}
A.~Hyv{\"a}rinen.
\newblock Fast and robust fixed-point algorithms for independent component
  analysis.
\newblock \emph{Neural Networks, IEEE Transactions on}, 10\penalty0
  (3):\penalty0 626--634, 1999.

\bibitem[Joel et~al.(2011)Joel, Caffo, van Zijl, and
  Pekar]{joel2011relationship}
S.E. Joel, B.S. Caffo, P.~van Zijl, and J.J. Pekar.
\newblock On the relationship between seed-based and ica-based measures of
  functional connectivity.
\newblock \emph{Magnetic Resonance in Medicine}, 66\penalty0 (3):\penalty0
  644--657, 2011.

\bibitem[McKeown et~al.(1997)McKeown, Makeig, Brown, Jung, Kindermann, Bell,
  and Sejnowski]{mckeown1997analysis}
M.J. McKeown, S.~Makeig, G.G. Brown, T.P. Jung, S.S. Kindermann, A.J. Bell, and
  T.J. Sejnowski.
\newblock Analysis of fmri data by blind separation into independent spatial
  components.
\newblock Technical report, DTIC Document, 1997.

\bibitem[Mennes et~al.(2012)Mennes, Biswal, Castellanos, and
  Milham]{mennes2012making}
M.~Mennes, B.~Biswal, F.X. Castellanos, and M.P. Milham.
\newblock Making data sharing work: The fcp/indi experience.
\newblock \emph{NeuroImage}, 2012.

\bibitem[Milham(2012)]{milham2012open}
M.P. Milham.
\newblock Open neuroscience solutions for the connectome-wide association era.
\newblock \emph{Neuron}, 73\penalty0 (2):\penalty0 214--218, 2012.

\bibitem[Oja et~al.(2001)Oja, Hyvarinen, and Karhunen]{oja2001independent}
E.~Oja, A.~Hyvarinen, and J.~Karhunen.
\newblock Independent component analysis, 2001.

\bibitem[Pillai et~al.(2002)]{pillai2002probability}
S.U. Pillai et~al.
\newblock \emph{Probability, Random Variables, and Stochastic Processes}.
\newblock Tata McGraw-Hill Education, 2002.

\bibitem[Rombouts et~al.(2005)Rombouts, Barkhof, Goekoop, Stam, and
  Scheltens]{rombouts2005altered}
S.A.R.B. Rombouts, F.~Barkhof, R.~Goekoop, C.J. Stam, and P.~Scheltens.
\newblock Altered resting state networks in mild cognitive impairment and mild
  alzheimer's disease: an fmri study.
\newblock \emph{Human brain mapping}, 26\penalty0 (4):\penalty0 231--239, 2005.

\bibitem[Sambataro et~al.(2010)Sambataro, Murty, Callicott, Tan, Das,
  Weinberger, and Mattay]{sambataro2010age}
F.~Sambataro, V.P. Murty, J.H. Callicott, H.Y. Tan, S.~Das, D.R. Weinberger,
  and V.S. Mattay.
\newblock Age-related alterations in default mode network: impact on working
  memory performance.
\newblock \emph{Neurobiology of aging}, 31\penalty0 (5):\penalty0 839, 2010.

\bibitem[Sorg et~al.(2007)Sorg, Riedl, M{\"u}hlau, Calhoun, Eichele, L{\"a}er,
  Drzezga, F{\"o}rstl, Kurz, Zimmer, et~al.]{sorg2007selective}
C.~Sorg, V.~Riedl, M.~M{\"u}hlau, V.D. Calhoun, T.~Eichele, L.~L{\"a}er,
  A.~Drzezga, H.~F{\"o}rstl, A.~Kurz, C.~Zimmer, et~al.
\newblock Selective changes of resting-state networks in individuals at risk
  for alzheimer's disease.
\newblock \emph{Proceedings of the National Academy of Sciences}, 104\penalty0
  (47):\penalty0 18760--18765, 2007.

\bibitem[Toro et~al.(2008)Toro, Fox, and Paus]{toro2008functional}
R.~Toro, P.T. Fox, and T.~Paus.
\newblock Functional coactivation map of the human brain.
\newblock \emph{Cerebral cortex}, 18\penalty0 (11):\penalty0 2553--2559, 2008.

\bibitem[Zuo et~al.(2010)Zuo, Kelly, Di~Martino, Mennes, Margulies, Bangaru,
  Grzadzinski, Evans, Zang, Castellanos, et~al.]{zuo2010growing}
X.N. Zuo, C.~Kelly, A.~Di~Martino, M.~Mennes, D.S. Margulies, S.~Bangaru,
  R.~Grzadzinski, A.C. Evans, Y.F. Zang, F.X. Castellanos, et~al.
\newblock Growing together and growing apart: regional and sex differences in
  the lifespan developmental trajectories of functional homotopy.
\newblock \emph{The Journal of Neuroscience}, 30\penalty0 (45):\penalty0
  15034--15043, 2010.

\end{thebibliography}
\bibliographystyle{plainnat}

\newpage
\section*{Appendix}
\subsection*{Proof of Theorem \ref{thm:same}}

\begin{proof}
 Since the true sources are truly homotopic and noise free,  we have:
\[
\bX_{i,1}= \bX_{i,2},
\]
where $\bX_{i,j}$, $j=1,2$, are the data of left and right hemispheres after dimension reduction by PCA. This is equivalent to 
\[
\bX_{1}= \bX_{2}.
\]
Without loss of generality, we assume  $\bX_{j}$ ($j=1,2$) are demeaned i.e. the row means of  $\bX_{j}$ are all 0.
Assume the singular value decomposition of the matrix $\bX_{1}/\sqrt{V}$ is 
\[
\bX_{1}/\sqrt{V}=\bU \Sigma \bV^T, 
\]
where $\bU$ is  of dimension $NQ\times (NQ-1)$ which consists of the left singular vectors of $\bX_{1}$, $\Sigma$ is a diagonal matrix  of dimension $(NQ-1)\times (NQ-1)$ with the singular values, and $\bV$ is of dimension $V\times (NQ-1)$ which consists of the right singular vectors. Both $\bU$ and $\bV$  are orthogonal. Note that, since $\bX$ is not full rank, only $NQ-1$ singular value are non-zero and thus only  $NQ-1$ singular vectors are estimated.
Then we have 
\[
\begin{aligned}
\bX/ \sqrt{V}&=[(\bX_1 / \sqrt{V})^T,(\bX_2 / \sqrt{V})^T]^T\\
 &=[ (\bU\Sigma\bV^T)^T,(\bU\Sigma\bV^T)^T ]^T\\
 &=[\bU^T,\bU^T]^T\Sigma \bV^T
\end{aligned}
\]
and
\[
\begin{aligned}
\tilde\bX/ \sqrt{V}&=[(\bX_1 / \sqrt{V}),(\bX_2 / \sqrt{V})]\\
 &=[ (\bU\Sigma\bV^T),(\bU\Sigma\bV^T)]\\
 &=\bU \Sigma [\bV^T,\bV^T].
\end{aligned}
\]
 Thus we can define 
\begin{equation}\label{equ:defk}
\bK:=\frac{1}{2}\Sigma^{-1} [\bU,\bU]\;\; \mbox{and}\;\; \tilde{\bK}:= \Sigma^{-1} \bU
\end{equation}
 to be the pre-whitening matrix that projects data onto the principal components:
\[
\bZ= \bK \bX\;\; \mbox{and}\;\;
\tilde{\bZ}= \tilde{\bK} \tilde{\bX},
\]
where $\bZ=\sqrt{V}\bV^T$ and
$\tilde{\bZ}=[\sqrt{V}\bV^T,\sqrt{V}\bV^T]$ are the whiten data for
H-gICA and GIFT respectively. Clearly, if we assume the random
variables $\bz$ and $\tilde{\bz}$ are taking values from the columns
of $\bZ$ and $\tilde{\bZ}$, then we will have:
\[
E[\bz \bz^T]=\bI \;\; \mbox{and} \;\;E[\tilde{\bz} \tilde{\bz}^T]=\bI.
\]

Suppose we have the same initial value for the vectors $\bw_p$ for all $p\in\{1,2,\cdots,Q\}$. No matter how the contrast function, $G$, is defined for each fixed $\bw$ we have:
\[
\begin{aligned}
E\{\bz g(\bw^T\bz)\}&=\frac{1}{V}\sum\limits_{v=1}^V \bZ(\cdot,v)g(\bw^T\bZ(\cdot,v))\\
&=\sum\limits_{v=1}^V \bV^T(\cdot,v)g(\bw^T\bV^T(\cdot,v))\\
&=\frac{1}{2}\left(\sum\limits_{v=1}^V \bV^T(\cdot,v) g(\bw^T\bV^T(\cdot,v))+\sum\limits_{v=1}^V \bV^T(\cdot,v) g(\bw^T\bV^T(\cdot,v))\right)\\
&=\frac{1}{2V}\sum\limits_{v=1}^{2V}\tilde{\bZ}(\cdot,v) g(\bw^T\tilde{\bZ}(\cdot,v))\\
&=E\{\tilde{\bz} g(\bw^T\tilde{\bz})\} \\
\end{aligned}
\]
where $g$ is defined as the derivative of the contrast function $G$. Similarly,
\[
 E\{g'(\bw^T\bz)\}\bw=E\{g'(\bw^T\tilde{\bz})\}\bw,
\]
where $g$ is the derivative of $G$. So we have:
\[
 E\{\bz g(\bw^T\bz)\}-E\{g'(\bw^T\bz)\}\bw=E\{\tilde{\bz} g(\bw^T\tilde{\bz})\}  -E\{g'(\bw^T\tilde{\bz})\}\bw
\]
Thus, following the FastICA algorithm \citep{hyvarinen1999fast}, it is easy to see that the estimates of $\bw_p$, $p=1,2,\cdots,m$, will be same for these two approaches. By the definition of $\bK$ and $\tilde{\bK}$ in \eqref{equ:defk}, we will have:
\[
\begin{aligned}
\hat{\bW}&=[\bw_1,\bw_2,\cdots,\bw_m]^T\bK\\
&=\frac{1}{2}[\bw_1,\bw_2,\cdots,\bw_m]^T [\tilde{\bK},\tilde{\bK}]\\
&=\frac{1}{2}[\hat{\tilde{\bW}},\hat{\tilde{\bW}}]\\
\end{aligned}
\]
and
\[
\begin{aligned}
\hat{\bS}&=\hat{\bW}\bX=\frac{1}{2}[\hat{\tilde{\bW}},\hat{\tilde{\bW}}][\bX_1^T,\bX_2^T]^T=\hat{\tilde{\bW}}\bX_1\\
\hat{\tilde{\bS}}&=\hat{\tilde{\bW}}\tilde{\bX}=\hat{\tilde{\bW}}[\tilde{\bX_1},\tilde{\bX_2}]=\hat{\tilde{\bW}}[\tilde{\bX_1},\tilde{\bX_1}]\\
\end{aligned}
\]
Thus follows the result of  Theorem \ref{thm:same}.
\end{proof}

\end{document}